\documentclass[twocolumn,aps,pra,10pt,showpacs,floatfix]{revtex4-2}
\usepackage{bm}
\usepackage{amsfonts}
\usepackage{amssymb}
\usepackage{amsmath}

\usepackage{graphicx}
\begin{document}

\title{Time crystals made of electron-positron pairs}
\author{Iwo Bialynicki-Birula}\email{birula@cft.edu.pl}
\affiliation{Center for Theoretical Physics, Polish Academy of Sciences\\
Aleja Lotnik\'ow 32/46, 02-668 Warsaw, Poland}
\author{Zofia Bialynicka-Birula}
\affiliation{Institute of Physics, Polish Academy of Sciences\\
Aleja Lotnik\'ow 32/46, 02-668 Warsaw, Poland}

\begin{abstract}
We show that in the greatly simplified model of mutually interacting electron-positron pairs and an electric field, time-crystal structures can spontaneously form. For a special choice of parameters we find periodic fluctuations of the pair number and the electric field.
\end{abstract}
\maketitle

\section{Introduction}

The concept of time crystals introduced originally by Wilczek \cite{wil,sw} has undergone various modifications \cite{ks} but has retained its basic property defined in a review paper \cite{sz} as  ``spontaneous breaking of time translation symmetry.'' In the present work we show that time crystals may appear in a relativistic system: the Dirac sea state of the electron-positron pairs coupled in a self-consistent way with the electric field. This is a continuation of our other attempt \cite{bb} to describe the appearance and disappearance of electron-positron pairs.

\section{Dirac-Heisenberg-Wigner function in QED}

Thirty years ago the analog of the Wigner function was introduced \cite{bgr,bdr} to describe the quantum state of the system of the Dirac particles in quantum electrodynamics. For historical reasons, it was named the Dirac-Heisenberg-Wigner (DHW) function. The DHW function is based on the expectation value, evaluated in a state $\vert\Phi\rangle$, of the commutator of the Dirac field operators $\Psi_\alpha(\vec r + {\vec s}/2,t)$ and $\Psi_\beta^\dagger({\vec r}-{\vec s}/2,t)$, ($\hbar=1 {\rm and} c=1$),
\begin{widetext}
\begin{align}\label{wig}
W_{\alpha \beta}(\vec r,\vec p,t)=-{\textstyle\frac{1}{2}} \int\! d^3s\exp(-i\vec p\cdot \vec s)\exp\left(-ie\int_{-1/2}^{1/2}\! d\lambda\,\vec s\cdot \vec A(\vec r +\lambda \vec s,t)\right)\langle \Phi\vert \left[\Psi_\alpha (\vec r + {\vec s}/2,t),\Psi_\beta^\dagger ({\vec r}-{\vec s}/2,t)\right] \vert\Phi\rangle.
\end{align}
\end{widetext}
The field operators are assumed to obey the Dirac equation in the presence of the electromagnetic field $A_\mu(\vec r,t)$. The second exponential factor in (\ref{wig}) makes the DHW function gauge invariant.

Here we study the interaction of the electron-positron system with the electromagnetic field with the use of the DHW function. This system was described in \cite{bgr} by the set of nonlinear integro-differential equations obeyed by the 16 components of the DHW function,
\begin{subequations}\label{set}
\begin{align}
{\sf D}_t f_0 + \vec{\sf D}\cdot {\vec g}_1 = 0,\label{f1}\\
{\sf D}_t f_1 + \vec{\sf D}\cdot {\vec g}_0 = -2mf_2,\label{f2}\\
{\sf D}_t f_2 + 2{\vec {\sf p}}\cdot {\vec g}_3 = 2mf_1,\label{f3}\\
{\sf D}_t f_3- 2{\vec {\sf p}}\cdot {\vec g}_2 = 0,\label{f4}\\
{\sf D}_t {\vec g}_0 + \vec{\sf D}f_1 - 2{\vec {\sf p}}\times {\vec g}_1 = 0,\label{f5}\\
{\sf D}_t {\vec g}_1 + \vec{\sf D}f_0- 2{\vec {\sf p}}\times {\vec g}_0 = -2m{\vec g}_2,\label{f6}\\
{\sf D}_t {\vec g}_2 + \vec{\sf D}\times {\vec g}_3 + 2{\vec {\sf p}} f_3 = 2m{\vec g}_1,\label{f7}\\
{\sf D}_t {\vec g}_3-\vec{\sf D}\times {\vec g}_2- 2{\vec {\sf p}} f_2 = 0,\label{f8}
\end{align}
\end{subequations}
where $m$ is the electron mass and
\begin{align}
{\sf D}_t = \partial_t + e\vec E(\vec r,t)\cdot \vec{\partial_p},\\
\vec{\sf D} = \nabla + e\vec B(\vec r,t) \times \vec{\partial_p}.
\label{local}
\end{align}
This set of equations is made self-consistent by including the Maxwell equations in which the coupling of the electromagnetic field to the electric current is described by the appropriate components of the DHW function,
\begin{subequations}\label{max}
\begin{align}
\partial_t\vec B =-\nabla \times \vec E,\label{m1}\\
\nabla \cdot \vec B = 0,\label{m2}\\
\partial_t\,\epsilon_0\vec E = \nabla \times \mu_0^{-1}\vec B -\vec j ,\label{m3}\\
\nabla \cdot \epsilon_0\vec E =\rho,\label{m4}
\end{align}
\end{subequations}
where
\begin{align}
\rho(\vec r,t) = e\int\!\frac{d^3p}{(2\pi)^3}\,f_0(\vec r,\vec p,t),\label{s1}\\
\vec j(\vec r,t) = e\int\!\frac{d^3p}{(2\pi)^3}\,{\vec g}_1(\vec r,\vec p,t).\label{s2}
\end{align}

Solving all these equations in the general case is a hopeless task. However, a greatly simplified model suffices to generate time-crystal structures. This model is defined as follows. We take Eqs.~(\ref{set}) and (\ref{max}) as the starting point. Next, we search for simple solutions of these equations disregarding the question of whether a corresponding quantum state $|\Phi\rangle$ exists. We hope that our model still describes certain aspects of the physical system under consideration. This hope is strengthened by the discovery of its rich and unexpected properties.

In our simple model, we assume that the magnetic field is absent and the electric field is unidirectional and homogeneous in space. Then, we may assume that the functions $f_i$ and $g_i$ do not depend on $\vec r$ and that all vectors point in the direction of the electric field. The coupling of electron-positron pairs to the electric field is through the electric current $\vec j$ which is proportional to ${\vec g}_1={\rm Tr}\{\gamma^0{\vec\gamma}\,\hat{W}\}$. In order to obtain a closed set of equations we also need ${\vec g}_2={\rm Tr}\{-i\gamma^0\vec\gamma\hat{W}\}$ which does not have a clear physical interpretation and $f_3={\rm Tr}\{\gamma^0\hat{W}\}$ related to the number of pairs $N_p$,
\begin{align}
f_3=\frac{2m}{\sqrt{m^2+{\vec p}^2}}(N_p-2).
\end{align}
Under our simplifying assumptions, the remaining components of $\hat{W}$ do not couple to our three chosen components and they can be set equal to zero.
The evolution equations for the DHW function are reduced to the set of partial differential equations for the relevant components of vectors,
\begin{subequations}
\label{dhw}
\begin{align}
D_t f_3(p,t)&=2p g_2(p,t),\\
D_t g_1(p,t)&=-2mg_2(p,t),\\
D_t g_2(p,t)&=2mg_1(p,t)-2pf_3(p,t),
\end{align}
\end{subequations}
where $D_t=\partial_t+E(t)\partial_p$.

In order to simplify the analysis, we convert this set of equations to the canonical form. This is achieved by replacing the independent variables $(p,t)$ by $(p_c-eA(t),t)$  in all functions, where $p_c$ is the canonical momentum. The functions dependent on $(p_c,t)$ will be distinguished by adding tildes,
\begin{align}
\tilde f_3(p_c,t)=f_3(p_c-e A(t),t),
\end{align}
\;etc. The two combined derivatives of the old functions are equal to the derivative with respect to $t$ of the new functions, for example, $D_t f_3(p_c-e A(t),t)=\partial_t \tilde f_3(p_c,t)$.

The set of differential equations for the new functions is,
\begin{subequations}
\label{bgr}
\begin{align}
\frac{d}{dt}\tilde f_3(p_c,t)&=2[p_c-eA(t)]\tilde g_2(p_c,t),\\
\frac{d}{dt}\tilde g_1(p_c,t)&=-2m\tilde g_2(p_c,t),\\
\frac{d}{dt}\tilde g_2(p_c,t)&=2m\tilde g_1(p_c,t)-2[p_c-eA(t)]\tilde f_3(p_c,t).
\end{align}
\end{subequations}
The derivatives with respect to momentum disappeared because the transition from the kinetic to the canonical momentum is mathematically equivalent to the method of characteristics. It is worth noting that this purely mathematical method based on ``the equivalence of a first order partial differential equation with a certain system of ordinary differential equations'' \cite{ch} acquired here a clear physical meaning of the transition from the kinetic to the canonical momentum.

We now complete the set of equations by adding the following two equations (remnants of Maxwell equations) which relate the vector potential $A(t)$ to the electric field $E(t)$ and couple the current to the electric field:
\begin{subequations}
\label{cpl}
\begin{align}
\frac{d}{dt}A(t)&=-E(t),\\
\frac{d}{dt}E(t)&=-\gamma\tilde g_1(p_c,t).
\end{align}
\end{subequations}
The constant $\gamma$ in Eq.~(12b) is needed to relate the phase-space distribution $\tilde g_1(p_c,t)$ to the electric current. This constant, however, plays no role in the analysis of the solutions of the evolution equations because it disappears when the evolution equations are expressed in terms of the dimensionless functions,
\begin{align}\label{id}
M_x&=\frac{e\gamma\tilde f_3}{2m},\;M_y=\frac{e\gamma\tilde g_1}{2m},\;M_z=\frac{e\gamma\tilde g_2}{2m},\nonumber\\X&=-\frac{eA}{m}+\frac{p_c}{m},\;P=\frac{eE}{m},\;\tau=m t.
\end{align}
The equations of motion in terms of these new functions are
\begin{subequations}
\label{new}
\begin{align}
\frac{d}{d\tau}M_x(t)&=2X(t)M_z(t),\\
\frac{d}{d\tau}M_y(t)&=-2M_z(t),\\
\frac{d}{d\tau}M_z(t)&=2M_y(t)-2X(t)M_x(t),\\
\frac{d}{d\tau}X(t)&=P(t),\\
\frac{d}{d\tau}P(t)&=-2M_y(t),
\end{align}
\end{subequations}
These equations have two constants of motion,
\begin{align}\label{ham}
H=P^2/2+2XM_y+2M_x
\end{align}
and
\begin{align}\label{mom}
{\mathcal M}^2=M_x^2+M_y^2+M_z^2.
\end{align}
They can be cast into the canonical Hamiltonian form $d/d\tau F=\{F,H\}$ with the following choice of the non-vanishing Poisson brackets,
\begin{align}\label{pb}
\{X,P\}=1,\,\{M_i,M_j\}=\varepsilon_{ijk}M_k.
\end{align}
The Hamiltonian cannot be interpreted as the energy because it is not positive definite; it can take any value.

\section{Time crystals}

The space of solutions of Eqs.~(\ref{new}) can be parameterized by $q$ and by the initial values of all dependent variables at $t=0$. For a random choice of parameters we obtain irregular orbits like the one shown in Fig.~1. For special choices of the initial values we obtain perfect time crystals.

\begin{figure}
\begin{center}
\includegraphics[width=8.5cm,
height=5cm]{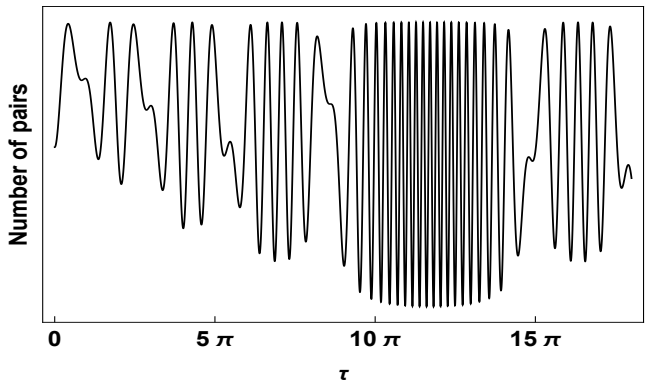}
\caption{Fluctuations in the number of pairs as a function of $\tau$ for a random choice of parameters.}
\end{center}
\end{figure}

\begin{figure}
\begin{center}
\includegraphics[width=8.5cm,
height=5cm]{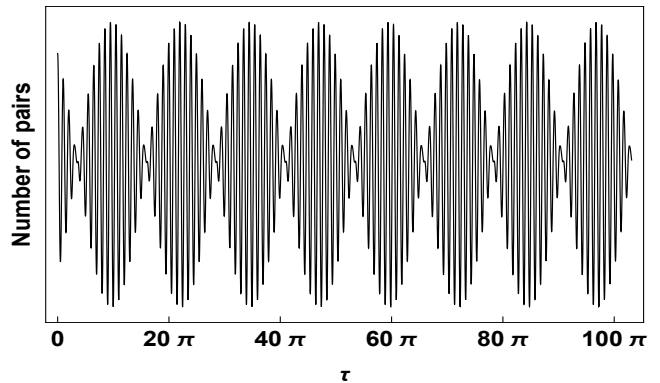}
\caption{Fluctuations in the number of pairs as a function of $\tau$ in the time crystal for $X(0)=0.2509$.}
\end{center}
\end{figure}

At first glance all time crystals have the same general form as shown in Fig.~2. They seem to have two periods of oscillations. The shorter period is equal to $\pi$ in our dimensionless units. This period is due to Zitterbewegung, {\em i.e.} to the oscillations of the quantum-mechanical phase of the wave function of the electron-positron pair with the frequency (in full form) $\omega=2c\sqrt{m^2c^2+p^2}/\hbar$. The modulation with the longer period $T$ describes a periodic bunching of pairs. Pairs appear and disappear forming a periodic structure, characteristic of {\em time crystal}. The changes of the number of pairs in time is accompanied, of course, by the periodic changes of the remaining variables.

\begin{figure}
\begin{center}
\includegraphics[width=7.52cm,
height=4.5cm]{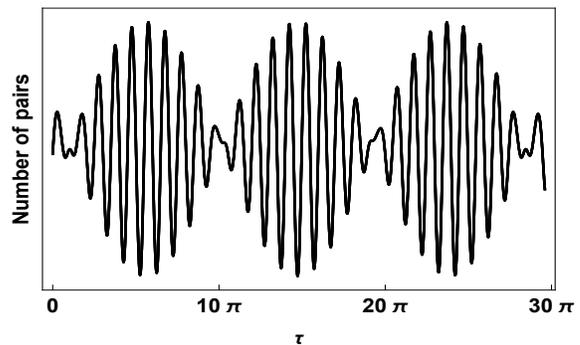}
\caption{The unit cell of the time crystal for
$X(0)=-0.12171$ and $T=27.43\pi$.}
\end{center}
\end{figure}

\begin{figure}
\begin{center}
\includegraphics[width=7.5cm,
height=4.5cm]{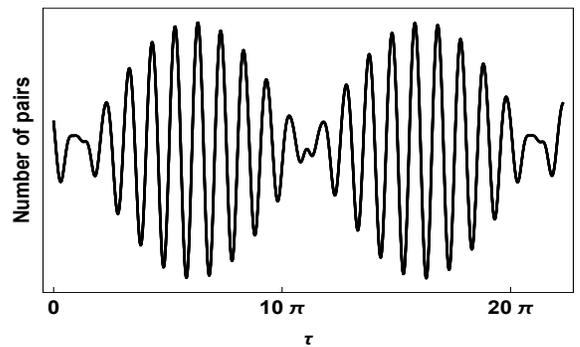}
\caption{The unit cell of the time crystal for $X(0)=-0.045$ and $T=20\pi$.}
\end{center}
\end{figure}

Closer scrutiny reveals a rich variety of time crystals. The unit cells of different time crystals have markedly different structures. We illustrate these differences in Figs.~3--6. Each figure represents four superimposed plots of the number of pairs $M_x(\tau),M_x(\tau+T),M_x(\tau+2T), {\rm and} M_x(\tau+3T)$. The fact that these plots differ less than the line thickness is a direct proof of periodicity. All figures were generated with the initial values,
\begin{align}\label{init}
M_x(0)=&0.009,\,M_y(0)=-0.027,\,M_z(0)=0,\nonumber\\
&P(0)=0.006.
\end{align}
The values of $X(0)$ were chosen by trial and error to obtain periodic structures. The intriguing feature shown in Figs.~3 and 4 is the substructure of the unit cells, which are made of very similar but not identical subunits.

\begin{figure}
\begin{center}
\includegraphics[width=7.5cm,
height=4.5cm]{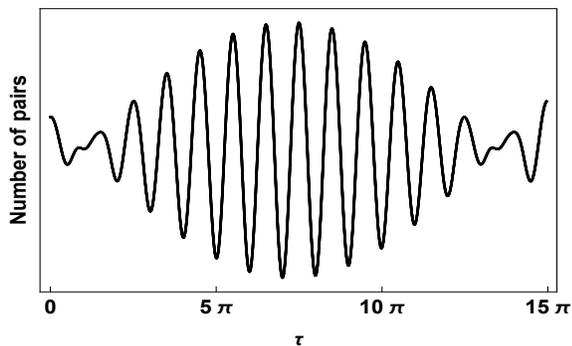}
\caption{Unit cell of the time crystal for $X(0)=0.0843$ and $T=12.465\pi$.}
\end{center}
\end{figure}

\begin{figure}
\begin{center}
\includegraphics[width=7.5cm,
height=4.5cm]{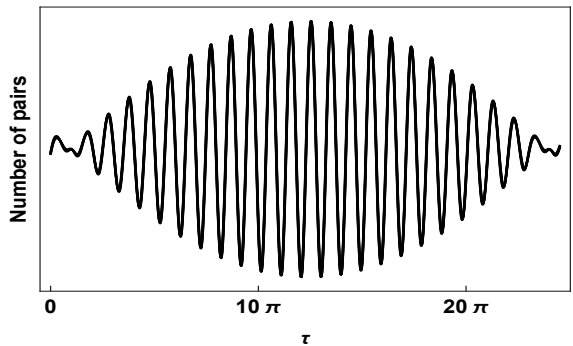}
\caption{Unit cell of the time crystal for $X(0)=0.2509$ and $T=23.025\pi$.}
\end{center}
\end{figure}

To end this section, we would like to add a few remarks about the chaotic behavior of the solutions of Eqs.~(\ref{new}). This set of equations has the same general structure (linear and quadratic terms on the right hand side) as the prominent equations (Lorentz, R\"ossler, and H\'enon-Heiles) which serve as workhorses in the analysis of chaotic behavior (cf., for example, \cite{bg}). The study of chaos in the system described by Eqs.~(\ref{new}) is worth a detailed analysis. Here we present only the results of our cursory investigation of this issue illustrated by the plots of the Poincar\'e maps for one of our periodic orbits and one chaotic orbit. The Poincar\'e maps in Figs.~7 and 8 were obtained by sampling the trajectory at times, when the value of the electric field $P(t)$ changes sign. The points represent the values of $M_x$ and $M_y$ at the sampling times.
\begin{figure}[t]
\begin{center}
\includegraphics[width=7cm,
height=4.5cm]{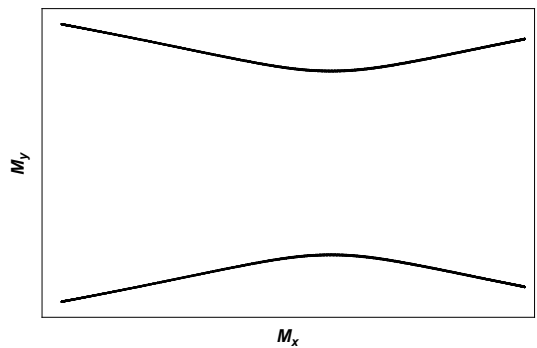}
\caption{Poincar\'e map for the periodic orbit with the same parameters as in Fig.~5.}
\end{center}
\end{figure}
\begin{figure}
\begin{center}
\includegraphics[width=7cm,
height=4.5cm]{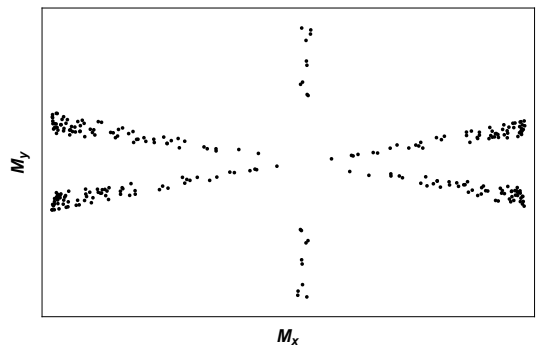}
\caption{Poincar\'e map for the chaotic orbit. This plot is obtained for the values of the parameters of Fig.~7 with the exception of the initial value of $M_z$, which was changed from 0 to 0.1743.}
\end{center}
\end{figure}

\section{The quantum Hamiltonian}

The full Hamiltonian structure of our dynamical system brings forth the question of its canonical quantization. The simplest version of the quantum Hamiltonian is obtained by replacing $X$ and $P$ by the operators $x$ and $-id/dx$ and $M_i$ by the Pauli matrices. In this way we reproduce the classical algebra of Poisson brackets by the quantum algebra of commutators. The quantum Hamiltonian becomes,
\begin{align}\label{qham}
{\hat H}=-\frac{1}{2}\frac{d^2}{dx^2}+2x\sigma_y+2\sigma_x.
\end{align}
\begin{figure}
\begin{center}
\includegraphics[width=7.5cm,
height=4.5cm]{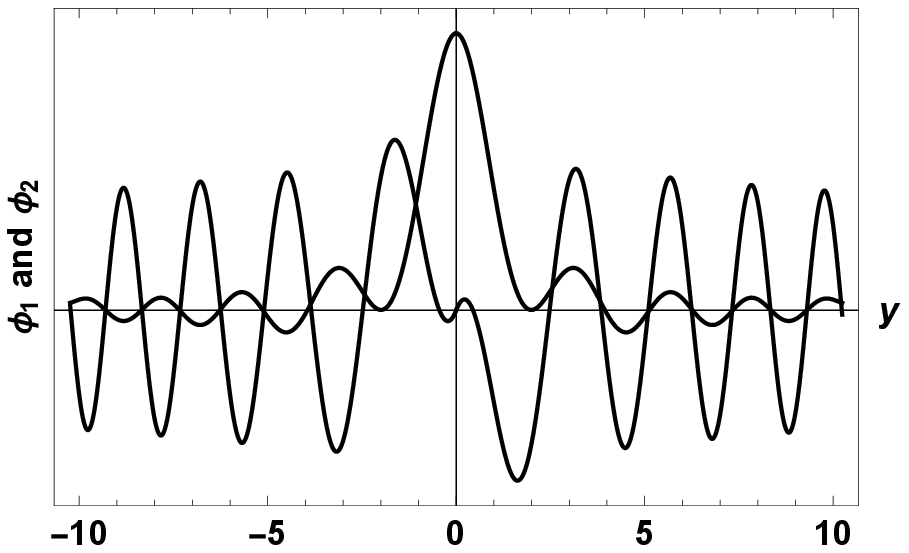}
\caption{Two components $\phi_1(y)$ and $\phi_2(y)$ of the eigenfunction of the quantum Hamiltonian (\ref{qham}) representing the first solution belonging to the eigenvalue $\mathcal{E}=2$. In order to generate regular functions we used at $y=0$ the values: $\phi_1=1,\,\phi_1'=0,\, \phi_2=0, {\rm and} \phi_2'=-0.354 651 985$.}
\end{center}
\end{figure}
\begin{figure}
\begin{center}
\includegraphics[width=7.5cm,
height=4.5cm]{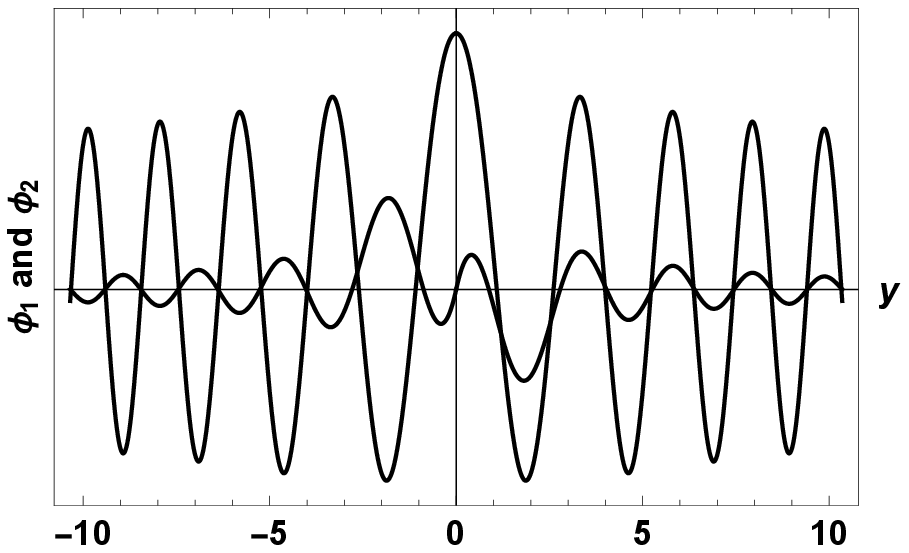}
\caption{Two components $\phi_1(y)$ and $\phi_2(y)$ of the eigenfunction of the quantum Hamiltonian (\ref{qham}) representing the second solution belonging to the eigenvalue $\mathcal{E}=2$. In order to generate regular functions we used at $y=0$ the values: $\phi_1=0,\,\phi_1'=-0.665 192 338,\, \phi_2=1,{\rm and} \phi_2'=0$.}
\end{center}
\end{figure}
The eigenvalue problem ${\hat H}\Psi=\mathcal{E}\Psi$ for this Hamiltonian is a set of two coupled second order equations for the two-component wave function $\{\psi_1,\psi_2\}$. After the rescaling $x=2^{-2/3}y$ and a convenient change of the basis,
\begin{align}\label{basis}
\psi_1(y)=\frac{1}{\sqrt{2}}\left[\phi_1(y)+i\phi_2(y)\right],\nonumber\\
\psi_2(y)=\frac{1}{\sqrt{2}}\left[\phi_2(y)+i\phi_1(y)\right],
\end{align}
we obtain the set of equations,
\begin{align}\label{eqsphi}
\left(-\frac{d^2}{dy^2}-\mathcal{E}/2^{1/3}+y\right)\phi_1(y)+2^{2/3}\phi_2(y)=0,\nonumber\\
\left(-\frac{d^2}{dy^2}-\mathcal{E}/2^{1/3}-y\right)\phi_2(y)+2^{2/3}\phi_1(y)=0.
\end{align}

\begin{figure}
\begin{center}
\includegraphics[width=7.5cm,
height=4.5cm]{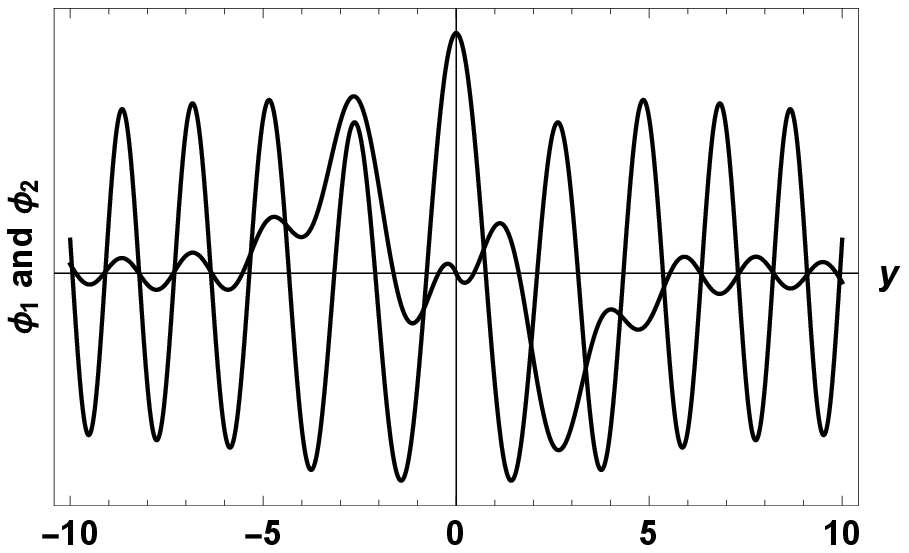}
\caption{Two components $\phi_1(y)$ and $\phi_2(y)$ of the eigenfunction of the quantum Hamiltonian (\ref{qham}) belonging to the eigenvalue $\mathcal{E}=5$. In order to generate regular functions we used at $y=0$ the values: $\phi_1=0,\,\phi_1'=-0.360 12,\, \phi_2=1,{\rm and} \phi_2'=0$.}
\end{center}
\end{figure}
We have not found analytic solutions but the application of a version of the shooting method (cf., for example, \cite{bck}) has enabled us to find the eigenfunctions. Our Hamiltonian has a continuous spectrum and, like its classical counterpart, is not bounded from below. The search for the eigenfunctions proceeded as follows. We have taken the eigenvalue $\mathcal{E}$ and searched for the initial values of the wave function which will produce a regular solution. This search was made simpler by bracketing the correct value, as in the standard shooting method. Since the normalization of the wave function is irrelevant, we took 1 as the initial value of one of the components and we searched for the value of the derivative of the other component which will produce regular functions. The two remaining initial values were set to 0. In this way, we generated the plot of the functions for positive values of $y$. The functions for negative values were generated for the same initial conditions at $y=0$. There is an expected double degeneracy for each value of the energy. Two solutions which belong to the same energy are shown in Figs. 9 and 10. We also calculated the two components of the eigenfunction for $\mathcal{E}=5$, shown in Fig.~11. In these plots, we scaled all eigenfunctions in the same way but, of course, the normalization of the eigenfunctions which belong to the continuous spectrum is arbitrary. All eigenfunctions seem to have a similar slow fall-off with increasing $|y|$ and the spacing between maxima decreases with the increase of the energy.

It remains an open question whether our periodic classical trajectories have their counterparts in the quantum case. They could appear as bound states embedded in the continuum but we found no trace of this.
\vspace{0.3cm}

\section{Conclusions}

We have shown that our model of electron-positron pairs coupled in a self-consistent way with the electric field, despite all the simplifications, exhibits rich properties seen at both the classical and the quantum levels. At the classical level we found periodic solutions of the differential equations that can be interpreted as time crystals made of electron-positron pairs. At the quantum level we found the eigenstates of the quantum Hamiltonian operator.

\end{document}